\documentstyle[prd,aps,epsfig,floats,tighten,preprint]{revtex}
\draft

\newcommand{\plb}[3]{Phys.\ Lett.\ {\bf B#1}, #3 (#2)}

\renewcommand{\apj}[3]{Astrophys.\ J.\ {\bf #1}, #3 (#2)}
\renewcommand{\prl}[3]{Phys.\ Rev.\ Lett. {\bf #1}, #3 (#2)}
\renewcommand{\prd}[3]{Phys.\ Rev.\ {\bf D#1}, #3 (#2)}

\begin{document}


\title{
\null
\vskip-6pt \hfill {\rm\small MCTP-01-56} \\
\vskip-6pt \hfill {\rm\small CERN-TH/2001-324}\\
\vskip-9pt~\\
Lensed Density Perturbations in Braneworlds:
An Alternative to Perturbations from Inflation}

\vspace{.5in}

\author{Daniel J. H. 
Chung$^{1,2}$ and Katherine Freese$^{1,3}$}

\address{
\vspace{.7cm}
$^1$ Michigan Center for Theoretical Physics,
Ann Arbor, MI 48109, USA\\
\vspace{.2cm}
$^2$ CERN Theory Division, CH-1211 Geneva 23, Switzerland\\
\vspace{.2cm}
$^3$ ISCAP, Columbia Astrophysics Laboratory,
550 W. 120th St.,
NY,NY 10027\\
}
\maketitle

\begin{abstract}
We consider a scenario in which our observable
universe is a 3-dimensional surface (3-brane) living in extra
dimensions with a warped geometry.
We show that ``lensed'' density perturbations from other branes serve
as possible seeds for structure formation on our observable brane
(without inflation),
and, in addition, provide constraints on
braneworld scenarios with warped bulk geometry.
Due to the warped bulk metric, any perturbation generated 
on one brane (or in the bulk matter) appears to 
an observer on a second brane to have a significantly different
amplitude.  We analyze lensed
perturbations in the Randall-Sundrum type scenarios and the ``shortcut
metric'' scenarios.
For Lorentz violating metrics in the bulk, 
we find the attractive possibility that large density 
fluctuations that are causally produced
elsewhere can lead to small density fluctuations on our brane on
superhorizon (acausal) length scales, as required by structure
formation.  Our most interesting result is that the
``shortcut metrics'' in which geodesics traverse the extra dimensions 
provide an alternative to inflation with two important features:
a possible solution to the horizon
problem and a mechanism to generate perturbations necessary for
structure formation.
\end{abstract}



\section{Introduction}

In the early universe, fluctuations in the matter density are thought to
grow via gravitational instability to give rise to the formation
of galaxies and other large scale structure.
Cosmic microwave background measurements \cite{cobe} constrain the amplitude
of the initial density perturbations to be
\begin{equation}
\label{eq:smallamp}
\delta\rho/\rho \sim 10^{-5} .
\end{equation}
In order to explain structure, these fluctuations must have existed on
large ``superhorizon'' length scales, larger than causally connected
regions in the early universe.  One mechanism to generate such
fluctuations is inflationary cosmology \cite{guth}, in which small
causally connected regions grow superluminally at a very early time to
encompass regions larger than our observable universe today.
Inflation is able to solve the horizon, flatness, and monopole
problems.  However, in most inflationary models, the smallness of the
perturbations in Eq.(\ref{eq:smallamp}) is difficult to maneuver.  The
magnitude of the perturbations is proportional to the ratio of the
height of the potential to the (width of the potential)$^4$
\cite{afg}.  Since the height and width are usually given by the same
mass scale, the smallness of this parameter is hard to
explain. Natural inflation \cite{natural} is a model in which a
pseudoNambuGoldstone boson provides an explanation for the two
different required mass scales.  However, in generic models, the small
number in Eq.(\ref{eq:smallamp}) is problematic.

One of the main points in this paper is to describe
a different generation mechanism
for density fluctuations. We consider a scenario in which
our observable universe is a 3-dimensional surface (3-brane) living
in higher dimensions.  In particular, we focus on a warped geometry
for the extra dimensions.  We work in 5 dimensions, but the
idea may be generalizable to higher dimensions.

Perhaps the most interesting application of lensed
perturbations is to ``shortcut metrics''
with geodesics that traverse the extra dimensions \cite{shortcut};
such metrics provide an alternative to inflation with two
interesting features: a possible solution to the horizon 
problem and generation of density fluctuations that can give rise
to structure formation.

For concreteness, we consider a metric of the form
\begin{equation}
\label{eq:cosmo}
ds^2 = - n^2(u,t) dt^2 + a^2(u,t) \delta_{ij} dx^i dx^j + du^2 ,
\end{equation}
where the coordinate $u$ parameterises the extra dimension.
Here, $a(u,t)$ is the scale factor of the 3-dimensional slices
parallel to our brane.  For most of the paper, we will treat the 
time dependence of the metric, at most, as a perturbation.
We will take the hypersurfaces defined by $u=0$ and $u=L$ to be
the world volumes of our observable universe brane and of a
``hidden sector'' brane.\footnote{By ``hidden sector'', we mean that the two 
branes do not communicate through any of the standard model fields
(we here use the language in which the graviton is not considered a standard
model field).}

The key here is to have a warped metric factor $a(u,t)$ that
changes rapidly as one leaves the brane along the $u$ direction,
leading to a change in comoving volume factor.  An example is
\begin{equation}
\label{eq:warp}
a(u,t) = e^{-bu}
\end{equation} 
for $0 \leq u \leq L$.  This has been considered previously in many
different physical contexts, e.g., by Randall-Sundrum \cite{RS1} and
\cite{RS2} in the context of the hierarchy problem and gravity
localization, and by Ref.~\cite{shortcut} in the context of the
horizon problem.  Consider a scenario in which our observable brane is
at $u=0$ and there is a second (hidden) brane at $u=L$.  With such a
metric, the 3-volume at $u=0$ is larger by a factor $e^{3bL}$ than the
volume at $u=L$ for a fixed coordinate volume; this change in comoving
3-volume is key to the lensing effect.  Now, imagine that density
fluctuations on very small scales but of very large amplitude
(e.g. ${\delta \rho_h \over \rho_h}|_L \sim 1$) are produced on this
hidden brane at $u=L$.  Here the subscript $h$ denotes the hidden
sector brane quantities.  For example, these perturbations might be
produced at some analog of a quark/hadron transition, which has has a
horizon size of a few kilometers and hence appears useless to
generating large scale density fluctuations. However, if such
perturbations are generated on the hidden brane, they will appear to
us to be on much larger scales and of much smaller amplitude.  For
very long wavelength modes, the amplitude is smaller by roughly the
ratio of brane volumes,
\begin{equation}
\underline{\delta \rho_h} \sim
e^{-c b L} {\delta \rho_h}, 
\end{equation}
where the underline refers to a projection onto our brane at $u=0$
(i.e. what the observers on the brane at $u=0$ gravitationally
perceive).\footnote{We do not define the brane tension to be part of
  these densities.  Furthermore, as we will explain later, for the
  Randall-Sundrum case this enhancement cannot be measured by a low
  energy observer, hence can be viewed as spurious.}  Here the factor
of $c$ in the exponent is a number of order 1 determined by the ratio
of volumes and the time measure which fixes the energy measurement.
In addition, in the case of Lorentz violating metrics (such as used in
\cite{shortcut}), perturbations produced on causal length scales are
blown up to much larger, acausal length scales on our brane.  Hence
our observable universe sees large scale, small amplitude fluctuations
such as are required for structure formation.

A word on notation:  Throughout the paper, underlined
quantities are those that have been projected elsewhere, usually
from the hidden brane onto the observable brane.

In a previous paper\cite{shortcut}, we proposed
a non-inflationary solution to the
cosmological horizon problem via shortcuts through the extra
dimensions. A signal travelling along an
extra-dimensional null geodesic may leave our 3-brane, travel into the
extra dimensions, and subsequently return to a different place on our
3-brane in a shorter time than the time a signal confined to our
3-brane would take.  Works related to this idea include
\cite{moffat,visser,kalbermann,ishihara,ishihara2,isometrybreakingpaper2,caldwell,davis,kalbermann2,kiritsis,gonzalezdiaz,csaki,nolte,quevedo,youm5,dubovsky,youm1,youm2,youm3,youm4,dvali}.
One of the metrics we considered had the warped form described in
Eq. (\ref{eq:warp}).  Hence, in such a shortcut scenario, it is
possible that small scale, large amplitude perturbations generated on
a hidden brane provide the necessary superhorizon small amplitude
fluctuations on our brane.  Any alternative to inflation should
address not only the horizon problem but also perturbation generation
as well as the flatness problem.  In fact, it is this question that
initiated the current investigation.  Here we propose and analyze the
features of a possible generation mechanism for the fluctuations in
the shortcut scenario.

We also wish to point out that lensed perturbations may be useful even
if our brane does have an inflationary period to solve the horizon and
the flatness problem.  If inflation cannot produce perturbations
required by Eq.(\ref{eq:smallamp}) (e.g., the height of the potential
is too low), lensed perturbations projected
onto our brane from elsewhere may be essential to the formation of
structure in the universe.

Our results apply equally to a single brane scenario.
For example, density fluctuations produced in the bulk will have the
same projection on our brane (as from a hidden brane) and may thus be
interesting for structure formation.

Randall and Sundrum first proposed a warped geometry (however without
shortcut geodesics) in order to suggest a solution to the hierarchy
problem.  We analyze lensed density perturbations in the RS scenario
as well.
For perturbations of a given amplitude on a hidden brane
with a Randall-Sundrum metric, we have found the size of the
perturbations seen by an observer on our brane and present the results
in Eqs.(\ref{eq:rsfinal}) and (\ref{eq:rslong}).  In the first paper
with warped geometry, RSI \cite{RS1}, the hidden brane is at $u=0$ and
the observable brane is at $u=L$; i.e., the locations of the hidden
and observable branes are exchanged from our previous discussion, such
that the warping goes in the opposite direction and the volume is
smaller on our brane.  In this situation, we find that homogeneous
(long wavelength) perturbations appear larger on our brane than where
they are generated.
However, from a low energy observer point of view, this projected
larger effective density is cloaked in the warp factor suppressed
volume factor such that the observed density is same as that seen on
the brane where the density locally resides.  Hence, the bound coming
from today's cosmological observations are same as that for the usual
dark matter.  

Note that one might expect an exact reversal of the lensing if the
observer were in the bulk instead of on a brane at
$u=0$: if large density perturbations produced elsewhere look small on
our brane, then one might expect small perturbations produced on our
brane to look large as one moves further and further away from our
brane. This situation applies to the second paper by Randall and
Sundrum, RSII \cite{RS2}.  Indeed, for large wavelength modes, this
reversal of lensing is true.  In that case, at a far enough distance,
one might fear that there are large back reaction corrections (for
related work, see \cite{GKR,gk,hebecker,chr,lup}).  However, as was
shown in Ref.\cite{GKR} and as is reviewed in the Appendix, the higher
momentum modes can still be suppressed with the result that there is
not necessarily a large back reaction.  The degree to which there is a
back reaction depends upon the degree of spatial localization of the
perturbation.  Roughly speaking, more extended sources lead to a
greater back reaction.

In the Discussion section we briefly address the flatness problem.
Although it seemed naively plausible that the difference in effective
mass density between lensed and intrinsic perturbations could be used
to ameliorate the flatness problem, we find that it does not help at
all.  We also mention a possible role of lensed perturbations in the
dark energy problem.  We speculate that large energy densities with a
negative equation of state on the hidden sector brane (or in the bulk)
may be projected to a small energy density on our brane and hence
provide an explanation of the current acceleration of the universe.

Finally, we would like to emphasise that although our calculations in
this paper bring out certain qualitative features of density
perturbations in braneworlds, details of their implications will
require more model dependent investigations by perturbing about
fully time dependent backgrounds.  Some work in braneworld density
perturbations in time dependent backgrounds can be
found in \cite{csaki2,bruck,brandenberger,koyama,langlois,kodama,veneziano} and
references therein. In this paper, for the most part, we do not consider 
the time evolution of the perturbations.

We begin by discussing lensed perturbations in a Randall Sundrum
metric in Section II.  The main aim of this section is to establish
notation and build intuition for the analysis of the shortcut metric
perturbations. 
For perturbations of a given amplitude on a hidden brane with
a Randall-Sundrum metric,
we have found the size of the perturbations seen by an observer
on our brane and present the results
in Eqs.(\ref{eq:rsfinal}) and (\ref{eq:rslong}).
Next we proceed to lensed perturbations in a
shortcut metric in Section III.  In Section IV, we discuss possible
generation mechanisms of perturbations on the hidden brane, comment on
the flatness problem, and close with a speculation of the use of lensing
on the dark energy problem.

\section{Density Perturbations in the Randall Sundrum metric}

In this section, we analyze the perturbations about the
Randall-Sundrum metric both from a 4D effective theory point of view
and the 5D point of view.  Our main objective is to establish notation
and build intuition for the analysis in the next section regarding the
perturbations about the shortcut metric.  In particular, although a 4D
effective theory description is sufficient to understand the
qualitative features of the perturbations in the Randall-Sundrum
scenario, since a covariant 4D effective theory description does not
exist in the shortcut scenario, we carry out both a 4D and a 5D
analysis in the Randall-Sundrum scenario as a warm up exercise to
compare to the 5D analysis of the shortcut scenario in the next
section.

We restrict our attention to 5 dimensions with the coordinate $ u $
parameterising the extra dimension.  In addition, for simplicity, let
the extra dimension be compactified on an $ S^{1}/Z_{2} $.  Consider 2
branes at orbifold fixed points, with brane 1 at $ u=L_1 $ and brane 2
at $ u=L_2 $. We would like to ask the question: given that there is a
density perturbation $ \delta \rho _{2} $ on brane 2, what is the
``effective density perturbation'' seen by the observer on brane 1?


We consider the perturbation about the AdS metric:
\begin{equation}
\label{eq:RSmetric}
ds^{2}= e^{-2bu}[\eta _{\mu \nu }+h_{\mu \nu }(x,u)]dx^{\mu }dx^{\nu }
+ du^{2}.
\end{equation}
Eq. (\ref{eq:cosmo}) reduces to this case when $a(u) = n(u) = e^{-2bu}$
and for linearised gravity,
$g_{\mu\nu} = \eta_{\mu\nu} + h_{\mu\nu}$, where $\eta_{\mu\nu}$ is the
metric of Minkowski space.
Here, Greek indices are four dimensional and range from
$\mu = 0,1,2,3$.
Then any matter on the branes is treated
as a perturbation to this background and serves as a source
for linearised gravity.  

A treatment of linearised gravity in the Randall-Sundrum background
has been worked out by \cite{GKR}.
Taking\footnote{We remind the reader
that underlined quantities indicate projected quantities from
one brane to another, and are not to be confused
with the overbar here.}
\begin{equation}
h_{\mu \nu }=\bar{h}_{\mu \nu }-\frac{\bar{h}}{d-2}\eta _{\mu \nu } 
\end{equation}
as the definition of $\bar{h}_{\mu\nu}$,
\begin{equation}
T_{\mu \nu }=S_{\mu \nu }(x)\delta (u-L_2) 
\end{equation}
as the 5 D stress tensor where $ S_{\mu \nu } $ is
the 4-dimensional stress tensor confined to the
brane 2 at $u=L_2$, and
\begin{equation}
T_{\mu u}\equiv 0 \,\,\, (e.g., \, T_{05}\equiv 0) ,
\end{equation}
they obtained
\begin{equation}
\label{eqorig}
\frac{1}{\sqrt{g}}\partial_{A}[\sqrt{g}g^{AB}
\partial_{B}]\bar{h}_{\mu \nu }(x,u)
\nonumber 
 =
 \frac{-e^{2bu}}{M^{3}}[S_{\mu \nu }-\frac{1}{6}
\eta _{\mu \nu }S^{\lambda }_{\lambda }]\delta (u-L_2)
-
\frac{1}{M^{3}}Q_{\mu \nu }
\end{equation}
where $ Q_{\mu \nu } $ contains the delocalized (not localized by a
delta function) stress energy related to the stabilisation of the
extra dimension, $M$ is the 5-dimensional Planck mass, and all Greek
indices are raised and lowered by $ \eta _{\mu \nu } $.

Let us pause here to explicitly display the warp factor that arises in
the case that the source $S_{\mu \nu}$ is a perfect fluid.  The five
dimensional stress tensor for a perfect fluid confined to the brane
located at $u=L_2$ can be written as
\begin{equation}
\label{eq:stress5}
T^{A}_{\, B}=[ p_{2}(\delta ^{A}_{\, \, B}-\delta
^{A5}\delta _{B5})-(p_{2}+\rho _{2})\delta ^{A0}\delta
_{B0}]\delta(u-L_2)
\end{equation} 
in the rest frame of the fluid, where $\rho_2$
and $p_2$ are energy density and pressure, respectively; 
i.e., $T^A_B = {\rm diag}(-\rho,p,p,p,0)$.  
Here Latin indices are 5-dimensional and range from $A = 0,1,2,3,4$.
These 5-dimensional indices are raised and lowered with the 
5-dimensional metric $g_{AB} = \rm{diag}(e^{-2bu} g_{\mu\nu},1)$. 
Now, stress tensor \( T_{\mu \nu } \) on brane 2
is the four-dimensional part of $T_{AB} = g_{AC}T^C_B$, i.e.,
\begin{eqnarray}
\label{eq:fourd}
T_{\mu \nu } & \equiv & e^{-2bu}\eta _{\mu A}T^{A}_{\, \, \nu }=\eta
_{\mu \alpha }T^{\alpha }_{\, \, \nu }\\ & = & e^{-2bu}[p_{2}\eta
_{\mu \nu }-(p_{2}+\rho _{2})\eta _{0\mu }\delta _{\nu 0}]
\delta(u-L_2)\\
&= &S_{\mu \nu} \delta(u-L_2)
\end{eqnarray}
Hence, note that $S_{\mu\nu}$ has
an additional warp factor compared to what one would naively
write down. 

\subsection{4D effective description}

Here we obtain a preview of part of our results for the
Randall-Sundrum case study by looking at the problem from a
four-dimensional point of view. We integrate over the 5th dimension
$u$ to obtain a 4D zero mode effective theory point of view.  In this
way one finds a result which holds in the long wavelength limit of the
more general study we perform below.  It is important to note that
from a 4D zero mode effective theory point of view, $p_2$ and $\rho_2$
in Eq.(\ref{eq:stress5}) cannot be measured.  Although this 4D
effective theory treatment is sufficient to accurately describe the
processes in the RSII case (as our fully 5D investigation will
confirm), since a 4D effective theory is not a good description of
gravity for the shortcut scenario, to be able to compare the shortcut
scenario with the Randall-Sundrum scenario, a 5D analysis of the
Randall-Sundrum scenario is required and will be carried out in the
subsequent subsection. 

Consider the action 
\begin{equation}
S=M^{3}\int d^{5}x\sqrt{g}R[g_{AB}]+
\int d^{5}x\sqrt{g}{\cal L}[g_{AB},\phi _{i}]\delta (u-L)+...
\end{equation}
where \( R \) is the Ricci scalar and \( {\cal L} \) is the 4D matter
Lagrangian. With the usual zero mode expansion ansatz 
\( ds^{2}=e^{-2bu}{\tilde{g}}_{\mu \nu }(x)dx^{\mu }dx^{\nu }+du^{2} \),
where we consider only the metric $\tilde{g}_{\mu\nu}(x)$ with no $u$-dependence,
one finds 
\begin{equation}
S\approx \frac{M^{3}}{2b}\int d^{4}x\sqrt{{\tilde{g}}}R_{4D}
[{\tilde{g}}_{\mu \nu }]+ S_M + ...
\end{equation}
where
\begin{equation}
S_M=\int d^{4}x\sqrt{\tilde{g}}e^{-4bL}
{\cal L}[e^{-2bL}{\tilde{g}}_{\mu \nu },\phi _{i}].
\end{equation}
This equation describes the 4D effective gravity interaction if the radion is
stabilised. In this case, the measurable stress tensor is the variation
with respect to \( \delta {\tilde{g}}_{\mu \nu } \). 
 Take the functional derivative with respect to \( {\tilde{g}}_{\mu \nu } \)
\begin{equation}
\frac{\delta S}{\delta {\tilde{g}}^{\mu \nu }}=
\frac{M^{3}}{2b}\sqrt{{\tilde{g}}}\left[ R_{4D\mu \nu }
-\frac{1}{2}{\tilde{g}}_{\mu \nu }R_{4D}\right] +
\frac{\delta S_{M}}{\delta {\tilde{g}}^{\mu \nu }}.
\end{equation}
 Hence, we arrive at the 0-0 component of the Einstein's equation\begin{equation}
\label{eq:4Dstresstriv}
R_{4D\, \, 0}^{\, \, \, 0}-\frac{1}{2}R_{4D}=
\frac{1}{M_{pl}^{2}}\frac{2}{\sqrt{{\tilde{g}}}}
\frac{\delta S_{M}}{\delta {\tilde{g}}^{\alpha 0}}{\tilde{g}}^{0\alpha }=
-\frac{\bar{\rho }^{(4D)}}{M_{pl}^{2}}
\end{equation}
where the second equality derives from the definition of 4D stress
tensor $\bar{\rho }_{2}^{(4D)}$
 that goes with the Einstein equations.

Now, let us vary the 5D metric to find an expression of \( \rho _{2} \)
as a functional of \( S_{M} \). \begin{eqnarray}
T^{0}_{\, 0}=-\rho _{2}\delta (u-L) & = & \frac{2}{\sqrt{g}}\frac{\delta S_{M}}{\delta g^{A0}}g^{0A}\delta (u-L)\\
 & = & \frac{2}{\sqrt{{\tilde{g}}}e^{-4bL}}\frac{\delta S_{M}}{\delta {\tilde{g}}^{A0}}{\tilde{g}}^{0A}\delta (u-L)
\end{eqnarray}
Comparing this with Eq. (\ref{eq:4Dstresstriv}), we see
\begin{equation}
\label{eq:comp}
\bar{\rho }^{(4D)}=e^{-4bL} \rho_{2}
\end{equation}
Hence the low energy quantity $\bar{\rho }_{2}^{(4D)}$
measured by a four dimensional
observer on brane 1 at $u=0$ due to energy momentum on brane 2
at $u=L$ is suppressed.

Thus, without a detailed calculation, we already see what the answer
to our question is in the zero momentum (long wavelength) limit. Since
the effective action for the matter confined to the brane at \( u=0 \)
is\[
\int d^{4}x\sqrt{{\tilde{g}}}{\cal L}({\tilde{g}}_{\mu \nu },\phi _{i})\]
which is unsuppressed, we see that the energy density projected onto
brane 1 from brane 2 will be suppressed by \( e^{-4bL} \) . Note
that this was already emphasised by Ref. \cite{csakigraesser}.

Note that there is no suppression in the other direction.
If there is a source at $u=0$, then the associated $\delta(u=0)$
gives rise to a factor $e^0=1$ so that there is no warp factor;
hence a source at $u=0$ appears unsuppressed to a source at $u=L$
i.e., 
\begin{equation}
\label{eq:nosuppress}
\bar{\rho }^{(4D)}= \rho(u=0) .
\end{equation}

In the following section, we perform a more complete general analysis in which
we study solutions to the equations of motion.  
Given a source on one brane, we find the effective density on another
brane that would be needed to reproduce the gravitational behaviour.  
Below both the source and effective density are 5-dimensional objects,
in that they are defined in a 5-dimensional stress tensor.
We will find that a source at location $L$ appears suppressed at $u=0$,
while a source at location $u=0$ appears enhanced at $u=L$.
However, as discussed in the previous paragraphs,
these 5-dimensional objects are not measurable observables.
One must still construct the 4-dimensional
object $\bar{\rho }_{2}^{(4D)}$ which appears in the 4D
effective action.  Then we confirm that
in the long wavelength limit one obtains
a suppressed effective 4D density for a source at $u=L$
at $u=0$ ; we also confirm that there is no suppression or
enhancement in the long wavelength limit in the observable density
for a source at $u=0$ on a brane at $u=L$.

\subsection{General Analysis}

Now we turn to a more general study in which we study
solutions to the equations of motion for density perturbations.
Rather than the coarse-graining approach of the previous paragraphs,
we perform a careful analysis of linearised gravity
(which includes Kaluza Klein modes).
Here the four-dimensional stress tensor $S_{\mu\nu}$ 
includes {\it all the matter} on brane 2, and 
is treated as a perturbation about the
much larger brane tension in the AdS background.
Hence, $S_{\mu \nu}$ 
includes not only the usual energy
density perturbation in cosmology (i.e. $\delta \rho$ in $\delta \rho/
\rho$) but also the energy density $\rho$ as well.

We keep terms only up to linear order in the expansion in $S_{\mu\nu}/B$,
where $B$ is brane tension.  Since we include both $\rho$ and
$\delta \rho$ in $S_{\mu\nu}$, this first order treatment
restricts the validity
of the analysis to the case where 
${\delta \rho \over B} > ({\rho \over B})^2$,
i.e.
\begin{equation}
\frac{\rho}{b^2 M_{pl}^2} < \frac{\delta \rho}{\rho}
\end{equation}
where $\mbox{brane tension} \sim b^2 M_{pl}^2 $.  
In the rest of this paper, we will almost always refer
to $S_{\mu \nu}$ as a density perturbation and will not distinguish between
$\rho$ background and $\delta \rho$ unless required.

In addition, the slice expansion in 3-spatial dimensions with local (in $u$)
expansion rate $\dot{a}/a(u)$ is encoded into $h_{ii} \sim 2
\dot{a}/a(u,t_{initial}) (t-t_{initial}) + ...$, and the equations are
only valid for time length scales smaller than
$\mbox{max}[a/\dot{a}]$.  Hence, any conclusions that one can draw
from our analysis regarding cosmological horizon physics is only
qualitative.  

Now we can begin to address the question:
given that there is a density perturbation $ \delta \rho
_{2} $ on brane 2, what is the ``effective density perturbation'' seen
by the observer on brane 1?
For generality, we take brane 1 to
be located at $u= L_1$ and brane 2 to be located at $L_2$. 
We define a 4 dimensional effective  $\underline{S}_{\mu\nu}$
as the 4-dimensional density perturbation
on brane 1 which reproduces the gravitational
effects generated by a density perturbation
$ S_{\mu \nu } $ on brane 2. Explicitly, we define it
as the source at $ u=L_1 $ such that 
\begin{equation}
\label{eq:effects}
\frac{1}{\sqrt{g}}\partial _{\alpha }
[\sqrt{g}g^{\alpha \beta }\partial _{\beta }]
\bar{h}_{1,\mu \nu }(x,u) =\frac{-e^{2bu}}{M^{3}}[\underline{S}_{\mu \nu }-
\frac{1}{6}\eta _{\mu \nu }\underline{S}^{\lambda }_{\lambda }]
\delta (u-L_1)-\frac{1}{M^{3}}Q_{\mu \nu }    
\end{equation}
where $ \bar{h}_{1,\mu \nu } $ matches $\bar{h}_{\mu \nu}$ of
Eq. (\ref{eqorig}) at $ u=0 $. Our aim will be to solve for $
\underline{S}_{\mu \nu } $.  
Note that although this projected stress
tensor is fictitious, it captures the gravitational information about
what the observer on brane 1 sees due to the stress tensor
perturbation on brane 2.  Also note that although this stress tensor
is intrinsically 4 dimensional since it is the response to the
variation of the induced metric, the induced metric describes the
geometry of a slice of a 5 dimensional space.  Hence, in that sense,
this stress energy tensor is really associated with 5 dimensional physics.

First look for the Green's function satisfying
\begin{equation}
\frac{1}{\sqrt{g}}\partial _{A}[\sqrt{g}g^{AB }
\partial _{B }]G(x,u;x_o,u_o)=\delta ^{(4)}(x-x_{0})\delta(u-u_o)
\end{equation}
or explicitly\begin{equation}
e^{2bu}[\partial _{t}^{2}G-\partial _{i}^{2}G]+4b\partial _{u}G-
\partial _{u}^{2}G=\delta ^{(4)}(x-x_{0})\delta(u-u_o) \, .
\end{equation}
Writing $ G $ as
\begin{equation}
G=\int \frac{d^{4}p}{(2\pi )^{4}}e^{ip\cdot (x-x_{0})}H(p,u,u_{0})
\end{equation}
 (where the dot product is only over the 0..3 coordinates $x$ and not
over the $u$ coordinate), we find
\begin{equation}
\label{eq:jumpeq}
H''-4bH'+e^{2bu}(p_{0}^{2}-\vec{p}^{\, 2})H=-\delta (u-u_{0}),
\end{equation}
where the prime denotes the derivative with respect to $ u $.
The integral of this equation yields the
discontinuity in $ H' $ at $ u=u_{0} $, i.e., the
jump condition at $u=u_0$.
Let us denote $ H_{>} $ to be the solution satisfying
the homogeneous equation for $ u>u_{0} $ and $ H_{<} $ to be the solution
for $ u<u_{0} $. Note that in the case that $ u_{0} $ approaches $ L_1 $,
it approaches $ L_1 $ from the right. Similarly, when $ u_{0} $ approaches
$ L_2 $, it approaches it from the left. Hence, we can write the
boundary conditions from  the jump equation as
\begin{equation}
\label{eq:bc1}
(H_{>}'-H_{<}')|_{u_{0}} = -1
\end{equation}
and
\begin{equation}
\label{eq:bc2}
H_{>}(u_{0}) = H_{<}(u_{0}).
\end{equation}
In the presence
of branes and neglecting back reaction effects due to sources on the branes,
we have the additional boundary conditions at the branes on orbifold 
fixed points,
\begin{equation}
\label{eq:bc3}
H_{<}'|_{u=L_1} = 0
\end{equation}
and
\begin{equation}
\label{eq:bc4}
H_{>}'|_{u=L_2} = 0 \, .
\end{equation}

This Green's function then determines the metric perturbation via
\begin{eqnarray}
\label{eq:metpert}
\bar{h}_{\mu \nu }(x,u=L_1) & = & \int d^{4}x_{0}du_{0}G(x,u=L_1;x_{0},u_{0})
\{\frac{-e^{2bu}}{M^{3}}[S_{\mu \nu }-\frac{1}{6}\eta _{\mu \nu }
S^{\lambda }_{\lambda }]\delta (u_0-L_2)-\frac{1}{M^{3}}Q_{\mu \nu }\}\\
 & = & \int d^{4}x_{0}du_{0}G(x,u=L_1;x_{0},u_{0})\{\frac{-e^{2bu}}
{M^{3}}[\underline{S}_{\mu \nu }-\frac{1}{6}\eta _{\mu \nu }
\underline{S}^{\lambda }_{\lambda }]\delta (u_0-L_1)-\frac{1}{M^{3}}Q_{\mu \nu }\}
\end{eqnarray}
Recall from Eq.(\ref{eq:fourd}) that $S_{\mu\nu} = e^{-2bu} 
{\rm diag}(-\rho,p,p,p)$.
Now we define
\begin{equation}
\underline{W}_{\mu \nu }\equiv \frac{-e^{2bL_1}}{M^{3}}
[\underline{S}_{\mu \nu }-\frac{1}{6}\eta _{\mu \nu }
\underline{S}^{\lambda }_{\lambda }] 
\end{equation}
and
\begin{equation}
\label{eq:wbar}
W_{\mu \nu }(x) =\frac{-e^{2bL_2}}{M^{3}}[S_{\mu \nu }(x)-\frac{1}{6}
\eta _{\mu \nu }S^{\lambda }_{\lambda }(x)] .    \end{equation}
 Using the Fourier convention
\begin{equation}
W_{\mu \nu }(x)=\int \frac{d^{4}p}{(2\pi )^{4}}\tilde{W}_{\mu \nu }(p)
e^{ip\cdot x}
\end{equation}
and similarly for $ \underline{W}_{\mu \nu } $, we can solve for
$ \underline{\tilde{W}}_{\mu \nu }(p) $ as \begin{equation}
\label{maineq}
\underline{\tilde{W}}_{\mu \nu }(p)=\tilde{W}_{\mu \nu
}(p)\frac{H(p,u=L_1,u_{0}=L_2)}{H(p,u=L_1,u_{0}=L_1)}. 
\end{equation}

Up to this point, the analysis has been rigorous.  To make the result
more transparent, we will make some simplifying assumptions which are
not necessarily all realistic.  However, the qualitative feature that
we wish to bring out does not seem to depend upon the unrealistic
features of the assumptions.  Let us assume that one can write the
projected perturbation stress tensor as a perfect fluid of the form
\begin{equation}
\underline{S}^{\mu \nu }=(\underline{p_2} \eta ^{\mu \nu }+(\underline{p_2} 
+\underline{\rho_2} )U^{\mu }U^{\nu })  ,
\end{equation}
 where the velocity
\begin{equation}
U^{\mu }\equiv \frac{dx^{\mu }}{d\tau }
\end{equation}
is that of a comoving fluid element and $\underline{p_2} $ and $
\underline{\rho_2} $ are scalars (pressure and energy density)
measured by an observer in a locally inertial frame that moves with
the fluid.\footnote{Note that these projected effectively ``Planck
brane'' densities do show up unscaled in the zero-mode effective
theory.  Hence, these can be measured by a low energy observer.}  We
are interested in a nearly static fluid configuration.  In that case $
U^{i} $ for $ i=1,2,3 $ can be neglected and the condition
\begin{equation}
U^{\mu }U_{\mu }=-1
\end{equation}
 gives
\begin{equation}
\label{eq:energy}
\underline{S}^{00}=\underline{\rho_2} e^{-2bL_1}
\end{equation}
and
\begin{equation}
\label{eq:trace}
\underline{S}^\lambda_{\,\,\,\lambda}= 
(-\underline{\rho_2} +3 \underline{p_2})e^{-2bL_1}.
\end{equation}
We remind the reader that underlined quantities refer to 
projections from brane 2 to brane 1.  Hence,
$\underline{\rho_2}$ is the effective perturbation density as seen on brane 1
due to a perturbation density produced on brane 2.

We seek a relation between the Fourier transform
of the density on brane 2,
\begin{equation}
{\hat{\rho}_2}(t,\vec{k}))=\int d^{3}xe^{-i\vec{k}\cdot 
\vec{x}}{\rho_2} (x) ,
\end{equation} 
and the Fourier transform of its projection onto brane 1,
\begin{equation}
\label{eq:ofinterest}
\underline{\hat{\rho}_2}(t,\vec{k})=\int d^{3}xe^{-i\vec{k}\cdot 
\vec{x}}\underline{\rho_2} (x) .
\end{equation}
We have
\begin{equation}
\label{eq:unFT}
\underline{\hat{\rho}_2}(t,\vec{k})=
\frac{-6M^3}{5+3\underline{w_2}}\int \frac{dk^{0}}{(2\pi )}
e^{-ik^{0}t}\underline{\tilde{W}}_{00}(k^{0},\vec{k})  .
\end{equation}
where we have defined $\underline{w_2}\equiv
\underline{p_2}/\underline{\rho_2}$ and assumed it to be independent
of time.\footnote{This will in general not be true.}
If the stress tensor on brane 2 can also be described by a perfect
fluid with an approximately constant equation of state of the form
$w_2=\underline{w_2}$, we  find
\begin{equation}
\underline{\hat{\rho}_2}(t,\vec{k})=  \int \frac{dk^0}{2 \pi}
e^{-i k^0 t} \frac{H(k,u=L_1,u_{0}=L_2)}{H(k,u=L_1,u_{0}=L_1)} 
\tilde{\rho}_2(k).
\end{equation}
Now, to simplify our results, we shall neglect any
$\dot{a}/a(u) \Delta t$ corrections where $\dot{a}/a(u)$ is the
spatial expansion rate of 3-slice at coordinate $u$ and $\Delta t$ is
the time difference with respect to some initial time.  We will also
assume for now that the density perturbation is static, i.e.,
\begin{equation}
\label{eq:static}
\tilde{\rho}_2(k)=\hat{\rho}_2(\vec{k})(2\pi )\delta (k^{0})    
\end{equation}
where the notation ``hat'' over a quantity refers to a Fourier transformed
static quantity (i.e., a function only of 3-vector $\vec{k}$).
 Hence, we have
\begin{equation}
\underline{\hat{\rho}_2}(\vec{k})
\approx  \frac{H(k^{0}=0,\vec{k},u=L_1,u_{0}=L_2)}
{H(k^{0}=0,\vec{k},u=L_1,u_{0}=L_1)}\hat{\rho }_{2}(\vec{k})
\label{eq:import}
\end{equation}
which gives the effective perturbation density on brane 1 due to
perturbation density on brane 2. In particular, if we live on brane 1,
then given a density perturbation generated on a hidden brane
identified as brane 2 at $u=L$, we can deduce its amplitude from the
point of view of our brane.

\subsection{Simple Example with Flat Metric}

To gain some intuition for the quantity in Eq.(\ref{eq:import}),
consider a noncompact flat 5D spacetime, $ds^2 = -dt^2 + d\vec{x}^2 +
du^2$, which is obtained from the Randall Sundrum metric of
Eq.(\ref{eq:RSmetric}) by setting $b=0$ and removing the boundaries.
The differential equation for the case $ p^{0}=0 $ is \begin{equation}
H''=\vec{p}^{\, 2}H.  \end{equation} We can impose the boundary
condition \begin{equation} H_{<}(u\rightarrow -\infty )=0 ,
\end{equation} \begin{equation} H_{>}(u\rightarrow \infty )=0 ,
\end{equation} and the jump condition $ (H_{>}'-H_{<}')|_{u_{0}}=-1 $
and $ H_{>}(u_{0})=H_{<}(u_{0}) $.  Suppose our brane (a probe brane)
is located at $u=0$ and the hidden brane at $u=L$.  We trivially
find\begin{equation}
\label{propagatortriv}
H(p^{0}=0,\vec{p},u=0,u_{0}=L)=\frac{e^{-|\vec{p}|L}}{2|\vec{p}|}.
\end{equation}
Thus we have    \begin{equation}
\underline{\hat{\rho}_2}(\vec{k})=
e^{-k L}\hat{\rho }_{2}(\vec{k})    .
\end{equation}
This equation gives us the size of density perturbations (see
Eq.(\ref{eq:ofinterest})) as seen on our brane at $u=0$ in terms of
density perturbations $\hat{\rho}_{2}(\vec{k})$ produced on the
$hidden$ brane at $u=L$.  One can see that $\hat{\underline{\rho_2
}}(\vec{k})$
does not get significant contribution
from wavelengths that are much shorter than the distance $ L $.
Also note that, for all wavelengths longer than $L$,
the exponential plays no role and the perturbations on our
brane would look just the same as the perturbation on the other brane.

\subsection{Projection of Perturbations in Randall Sundrum Metric}

Let us now return to the RS-type scenario (AdS embedding).
Using the boundary conditions in Eq.(\ref{eq:bc1}-\ref{eq:bc4})
in Eq.(\ref{eq:jumpeq}), we find
solutions for $H(k,u,u_o)$.  Using these solutions in
Eq.(\ref{eq:import}), we find 
the amplitude $\underline{\hat{\rho}_2 }(\vec{p})$
of lensed density fluctuations projected onto brane 1, due to
the density fluctuations on brane 2, $\hat{\rho}_{2}(\vec{p})$.

\subsubsection{Observable brane as brane 1 at $u=0$:}

First, suppose that the hidden sector brane is brane 2 at $u=L$ while
the observable brane is brane 1 at $u=0$ (see for example,
Ref.~\cite{ced}).  Then we find the amplitude $\underline{\hat{\rho}_2
  }(\vec{p})$ of lensed density fluctuations projected onto brane 1,
due to the density fluctuations on brane 2, $\hat{\rho}_{2}(\vec{p})$:
\begin{equation}
\label{eq:rsfinal}
\underline{\hat{\rho}_2}(\vec{p})=
\frac{e^{-3bL}}{\frac{p}{b}[I_{2}(\frac{p}{b})K_{1}
(\frac{p}{b}e^{bL})+I_{1}(\frac{p}{b}e^{bL})K_{2}
(\frac{p}{b})]}\hat{\rho }_{2}(\vec{p})
\end{equation}
where $ I_{\nu }(z) $ and $ K_{\nu }(z) $ are modified Bessel function
of the first and second kind. In the limit that $ (p/b) e^{bL} \ll 1 $, we
have
\begin{equation}
\label{eq:rslong}
\lim _{p\rightarrow 0}
\underline{\hat{\rho}_2 }(\vec{p})
=e^{-4bL}\hat{\rho }_{2}(\vec{p})
\end{equation}
while in the limit that $ (p/b) e^{bL}\gg 1  $, we have
\begin{equation}
\label{eq:rsshort}
\underline{\hat{\rho}_2 }(\vec{p})
\sim 2e^{-5bL/2}e^{\frac{-p}{b}(e^{bL}-1)}\hat{\rho }_{2}(\vec{p}).    
\end{equation}

Eqs. (\ref{eq:rsfinal},\ref{eq:rslong},\ref{eq:rsshort}) are key
results which answer our original question for the Randall-Sundrum
case: given a (Fourier transformed, static) density perturbation
$\hat{\rho}_2(\vec{p})$ on brane 2, these equations tell us the
amplitude $\underline{\hat{\rho}_2}(\vec{p})$ of the lensed
perturbations as seen by brane 1.

We have obtained the following
important result: perturbations intrinsic to the hidden brane appear to
be reduced in amplitude by an exponential factor.  One can understand
the $ e^{-4bL} $ dilution in the long wavelength limit as
follows. First, the density simply scales as the spatial volume.
Since the volume at $u=0$ is larger by $e^{3bL}$ on our brane, the
density is reduced by the inverse of this factor.
Second, in addition to the spatial volume dilution, the time interval
is increased in projecting onto the Planck brane.  Hence, the energy
is redshifted by a factor $e^{bL}$ such that the energy density is
reduced by this additional factor, giving a total of $ e^{-4bL} $ for
the potential.  As a result, perturbations as seen by our
brane are reduced by an overall factor of $e^{-4bL}$.  

In Section IIA, we discussed that $\rho_2$ cannot be measured from low
energy measurements (measurements that do not resolve the bulk); one
must derive an effective 4D observable describing a low energy
observer's point of view.  What we have just described from a 5D point
of view is precisely this 4D observable, and it is in agreement with
section IIA in the long wavelength limit.

This result leads to the interesting possibility that large amplitude
perturbations produced on brane 2 may still satisfy Eq. (1) and
serve as a mechanism for generation of density perturbations on our brane.

One might hope to solve the flatness problem due to this large
suppression factor.  If the $e^{-4bL}$ suppression is
sufficiently strong, one might naively suspect that lensed perturbations
of very small amplitude might be able
to smooth out initially large perturbations on brane 1. However, as
we will discuss later, this idea fails.

Because of reciprocity, small fluctuations on brane 1 can imprint
large perturbations elsewhere.  Even in scenarios without a second
brane, as in RSII, one might fear that small fluctuations on our brane
will appear large as one moves farther and farther from our brane and
give rise to back reactions.  (For related work, see
\cite{GKR,gk,chr,lup}.)  However, as was shown in Ref.\cite{GKR} and
as is reviewed in the Appendix, there is not necessarily a large back
reaction because the higher momentum modes can still be suppressed.
The degree to which there is a back reaction depends upon the degree
of spatial localization of the perturbation.  Roughly speaking, the
less localization leads to a greater back reaction.  
Further discussions of these issues can be found in the Appendix.

Our primary result for this section is that large amplitude
perturbations produced on brane 2 (from a 5 dimensional point of view)
may look suppressed from the point of view of brane 1, such that they
satisfy Eq.(1) and hence can provide seeds for structure formation in
our observable universe.
 
\subsubsection{Our observable brane as brane 2 at $u=L$:}

The RSI scenario attempts to solve the hierarchy problem with an
inverse warping of the one we have been studying above:  hence
we will now consider the observable
sector at $u=L$ and the hidden sector at $u=0$.  Any long wavelength
density perturbations produced in the extra dimensions may be
exponentially enhanced from the point of view of a high energy
observer on our brane.  
We have
\begin{equation}
{H(k^0=0,\vec{k},u=L,u_0=0) \over
H(k^0=0,\vec{k},u=L,u_0=L)} =
{e^{2bL} \over {p\over b}[I_2({p\over b} e^{bL}) K_1({p\over b}) 
+I_1({p \over b}) K_2({p\over b} e^{bL})]} .
\label{eq:reverse}
\end{equation}
In the limit
$(p/b)e^{bL} \ll 1$, this goes to
\begin{equation}
\lim _{p\rightarrow 0} 
{H(k^0=0,\vec{k},u=L,u_0=0) \over
H(k^0=0,\vec{k},u=L,u_0=L)} =
e^{4bL}.
\end{equation}
Hence we find in the long wavelength limit
\begin{equation}
\label{eq:enhanced}
\underline{\hat{\rho_1}}(\vec{p}) = e^{4bL} \hat{\rho}_1(\vec{p}) 
\end{equation}
where consistently as before $\underline{\hat{\rho_1}}$ is the projected
density onto $u=L$ due to a source at $u=0$ (at brane 1).

However, as noted in Section IIA, from a low energy
observer's point of view (with the radion stabilised) who sees only
the massless KK mode, the enhancement is invisible: the enhancement of
the effective density is cancelled by the warp factor such that the
effective gravity sees exactly the original energy density of the
Planck brane.  (This has already been noted by
Ref.\cite{csakigraesser}.)
One can see this from Eq.(\ref{eq:nosuppress}):
\begin{equation}
\label{eq:useall}
\bar{\rho }^{(4D)} \sim \hat{\rho}_1
=e^{-4bL} (\hat{\rho}_1 e^{4bL})
=e^{-4bL} (\underline{\hat{\rho}}_1) .
\end{equation}
Here, we have used Eq.(\ref{eq:enhanced}) in the last equality.
Indeed the first and last expressions in Eq.(\ref{eq:useall})
satisfy Eq.(\ref{eq:comp}) for the four dimensional effect
of a source at $u=L$ (although $\underline{\hat{\rho}}_1$ here really is an
effective source).

The matter on the hidden sector brane looks like dark matter to us,
which can be constrained from observations.  Even though from a low
energy observer's point of view, the lensed factor is cancelled, given
that the natural scale on the Planck brane is the Planck scale,
mechanisms suppressing Planckian densities are necessary to have a
phenomenologically acceptable model.  Here we have a hierarchy problem:
we have no explanation for the fact that 
the localized density on our observable brane is
exponentially smaller than the localized density on the hidden sector
Planck brane.  We refer the reader to Ref.\cite{csakigraesser} for more
discussion on this.

\subsubsection{Scale of Perturbations}

We return to the scenario with our observable brane at $u=0$ (brane 1).
It may seem that, even without Lorentz violation in the
bulk (see e.g.\cite{shortcut,isometrybreakingpaper2}), the
perturbations on brane 2 can lead to apparently acausal perturbations to be
projected onto our brane.  Because of the relationship    
\begin{equation}
\label{eq:mom}
p_{\textrm{phys }2}=e^{bL}p_{\textrm{phys }1}    \end{equation}
for a fixed coordinate momentum vector, short wavelength perturbations on brane 2 can look like long wavelength perturbations
on brane 1.  However, for a cosmological scale factor 
independent of $u$, i.e. 
\begin{equation}
ds^2 = e^{-2b u}(- dt^2 + a^2(t) d\vec{x}^2) + du^2
\end{equation}
the physical horizon size is also larger by the same factor
on brane 1. Hence perturbations generated on causal scales (smaller
than the horizon) on brane 2 remain on causal scales on brane 1.
This fact is a reflection of the fact that this metric
is conformal to Friedmann Robertson Walker with an extra 
dimension\footnote{There is no acausality in an FRW metric.}
($-dt^2 + a^2 d\vec{x}^2 + dz^2$), and that conformal transformations
leave invariant the causal structure of the metric.

The size of the comoving horizon is determined by setting
$ds^2 = 0$ such that we find $d_{comoving} = \int_0^t {dt' \over a(t')}$.
The size of the physical horizon is then
\begin{equation}
d_H(u,t) = e^{-bu} a(t) \int_0^t {dt' \over a(t')} .
\end{equation}
Thus on brane 1 at $u=0$, the horizon size is 
$d_H(u=0,t) = a(t) \int_0^t {dt' \over a(t')}$,
while on brane 2 at $u=L$, the horizon size is
$d_H(u=L,t) = e^{-bL} a(t) \int_0^t {dt' \over a(t')}$.
Indeed the physical horizon is larger on brane 1 by 
\begin{equation}
d_H(u=0,t) = e^{bL} d_H(u=L,t) ,
\end{equation}
the same factor as in Eq.(\ref{eq:mom}).  Both the wavelength of the
perturbation and the
horizon are shifted by the same amount.
Hence, with any metrics which maintain
$SO(1,3)$ isometry (neglecting the time dependence of the
scale factor), including the RS metrics,
one cannot induce perturbations on scales beyond the
horizon length from perturbations below the horizon length.  
In the next section we consider a scenario in which $SO(1,3)$ is
violated by the warping and hence acausal perturbations can
be generated.

Finally, note that from Eq.~(\ref{eq:reverse}) for the situation where we
live on brane 2 and the 
perturbation is on brane 1 (e.g. RSI), 
the brane 1 perturbation wavelength must be longer than
$e^{bL}/b$ to show up as unsuppressed perturbation on our brane with a
wavelength longer than $1/b$.

\section{Density Perturbations in Scenarios with ``Shortcut Geodesics''}

Ref.~\cite{shortcut} proposed a non-inflationary solution
to the cosmological horizon problem in braneworlds.  A signal
travelling along an extra-dimensional null geodesic may leave our
three-brane, travel into the extra dimensions, and subsequently return
to a different place on our three-brane in a shorter time than the
time a signal confined to our three-brane would take.  Hence, these
geodesics may connect distant points which would otherwise be
``outside'' the four dimensional horizon (points not in causal contact
with one another).  Such shortcut metrics break $SO(1,3)$
isometry along the coordinate directions parallel to the 3-brane;
i.e. these metrics explicitly break Lorentz isometry.

Here we discuss the possibility that perturbations created elsewhere
on small length scales may be lensed to appear on our brane
on apparently acausal superhorizon length scales.  In other words,
not only may there be an explanation of large scale smoothness
via these shortcut metrics, but also an explanation of perturbations
on superhorizon scales.

Following \cite{shortcut}, we consider the Lorentz violating spacetime
(without $ SO(1,3) $ isometry),
\begin{equation}
ds^2= -dt^2 + e^{-2bu}  d{\vec{x}}^2  +du^2.
\end{equation}
We perturb about this background spacetime as
\begin{equation}
ds^{2}=du^{2}-dt^{2}[1+h_{00}(x,u)]+e^{-2bu}
[\delta _{ij}+h_{ij}(x,u)]dx^{i}dx^{j} + 2 e^{-b u} h_{0 i} dx^i dt.
\end{equation}
If we assume that the gravity interaction is point-like, the equation of motion
can be read off from Ref. \cite{misner} and is
\begin{equation}
\frac{1}{\sqrt{g}}\partial _{A}
[\sqrt{g}g^{AB}\partial _{B}]h_{00}\sim \frac{-A}{M^{3}}
\rho_2 \delta (u-L)
\end{equation}
where $ A $ is a constant coefficient of order 1 which depends on the
equation of state.  We can use the same Green's function formalism as
before. We find the general homogeneous solution with $ p^{0}=0 $ to
be \begin{equation} H(p,u,L)=[c_{1}-c_{2}z(u)]\cosh
[z(u)]+[c_{2}-c_{1}z(u)]\sinh [z(u)] \end{equation} where
\begin{equation} z(u)=\frac{p}{b}e^{bu} \end{equation} with $ p\equiv
|\vec{p}| $, and $c_1$ and $c_2$ are constants to be specified by the
boundary conditions. Imposing the boundary conditions, we
have\begin{eqnarray*} H_{<}(p^{0}=0,\vec{p},u,u_{0}=u_{0}) & = &
\frac{-e^{-3bu_{0}}}{p^{3}\sinh
[\frac{p}{b}(1-e^{bL})]}(pe^{bu_{0}}\cosh
[\frac{p}{b}(e^{bL}-e^{bu_{0}})]+b\sinh
[\frac{p}{b}(e^{bL}-e^{bu_{0}})])\times \\ & & (pe^{bu}\cosh
[\frac{p}{b}(1-e^{bu})]+b\sinh [\frac{p}{b}(1-e^{bu})])
\end{eqnarray*}
 and \begin{eqnarray*}
H_{>}(p^{0}=0,\vec{p},u,u_{0}=u_{0}) & = & \frac{-e^{-3bu_{0}}}{p^{3}\sinh [\frac{p}{b}(1-e^{bL})]}(pe^{bu}\cosh [\frac{p}{b}(e^{bL}-e^{bu})]+b\sinh [\frac{p}{b}(e^{bL}-e^{bu})])\times \\
 &  & (pe^{bu_{0}}\cosh [\frac{p}{b}(1-e^{bu_{0}})]+b\sinh [\frac{p}{b}(1-e^{bu_{0}})]).
\end{eqnarray*}
These expressions can be used as before to yield
\begin{equation}
\underline{\hat{\rho}_2}(\vec{p})=
\frac{e^{-2bL}}{\cosh [\frac{p}{b}(e^{bL}-1)]+
\frac{1}{(p/b)}\sinh [\frac{p}{b}(e^{bL}-1)]}\hat{\rho}_2(\vec{p})
\end{equation}
Again, underlined quantities refer to the fictitious effective
quantities on brane 1.
Given a density perturbation $\hat{\rho }_2(\vec{p})$
on brane 2,
$\underline{\hat{\rho}_2}(\vec{p})$ is the lensed perturbation
seen by observers on brane 1.

Note that the perturbation is exponentially suppressed.
In the limit that $ (p/b) e^{bL}\ll 1 $,
we have     \begin{equation}
\lim _{p\rightarrow 0}\underline{\hat{\rho}_2}(\vec{p})
\sim e^{-3bL}\hat{\rho_2 }(\vec{p})    \end{equation}
while in the limit that $ (p/b) e^{b L}\gg 1  $, we have
\begin{equation}
\underline{\hat{\rho }_2}(\vec{p})\sim
e^{-2bL}e^{-\frac{p}{b}(e^{bL}-1)}\hat{\rho}_2(\vec{p})
\end{equation}
The $ e^{-3bL} $ suppression for
the long wavelength case is easy to understand as before. Any energy density
on the Planck brane is diluted by the volume factor.

Let us now discuss the scale of the perturbations for the shortcut
 metric background.
 Since on brane 2 the momentum vector is enhanced by
\begin{equation}
p_{\textrm{phys }2}=e^{bL}p_{\textrm{phys }1},
\end{equation}
 short wavelength perturbations on brane 2 can look like long
wavelength perturbations on brane 1.
Here the size of the comoving horizon, obtained by setting $ds^2 = 0$, 
is $d_{comoving} = e^{bu} \int{dt }$.  Then
the size of the real (proper) horizon is $d_H= e^{-bu} d_{comoving}$,
such that $d_H = \int{dt }$ and is independent of $u$.
Hence perturbations created on causal length scales on brane 2
can appear on much larger superhorizon length scales on brane 1.

Suppose brane 2 corresponds to the hidden sector brane while
brane 1 corresponds to the observable sector brane.  Small scale
density fluctuations produced elsewhere in the universe appear on our
brane as large scale fluctuations of smaller amplitude. This scenario may
thus both potentially explain the large scale smoothness of the
universe (the horizon problem) as well as explain the superhorizon
small density perturbations required to provide seeds for large scale
structure.  Of course, there still remains a significant challenge in
identifying a physical system that generates the appropriate $SO(1,3)$
isometry breaking warped background.

\section{Discussion}

We have found that perturbations may be lensed by a warped geometry,
such that both their scale and amplitude may look different to an
observer on our brane than where they were originally produced.  We
have found that fluctuations can look large or small because of the
volume factors.  For example, if the volume on our brane is larger
than on a hidden brane for a fixed coordinate volume, large
perturbations on a hidden brane may appear to be small perturbations
on our observable brane due to a warping of the geometry.  The most
interesting application of lensed perturbations is to ``shortcut
metrics'' \cite{shortcut} in which geodesics traverse the extra
dimensions.  We found that large amplitude perturbations of small
wavelength on one brane can be lensed to be small amplitude
perturbations of superhorizon length scales here as is required for
structure formation.  Hence shortcut metrics provide an alternative to
inflation with two important features: a possible solution to the
horizon problem and a mechanism to generate perturbations necessary
for structure formation.

Suppose the perturbations on large scale are negligible for
matter confined to our brane, i.e., $\delta \rho_0=0$, where the
subscript $0$ denotes our observable brane confined fields. The
quantity of interest is
\begin{equation}
\frac{\delta \rho_{tot}}{\rho_{tot}} = \frac{\delta
\underline{\rho}_h}{\rho_0 + \underline{\rho}_h} \, ,
\end{equation} where $\rho_{tot}$ includes the
energy density $\rho_0$ intrinsic to our observable universe, and the
underline as before indicates the projection onto our brane.  If
$\underline{\rho}_h \ll \rho_0$, then the
amplitude of $\delta \rho_{tot}/\rho_{tot}$ can be very much
smaller than $\delta \rho_h/\rho_h$, the original amplitude of the
perturbations on the hidden brane (subscript h) where they were produced.

In addition, for the case of Lorentz violating metrics
such as the shortcut metrics, there is an apparent acausal behaviour of the
perturbation.  As the scale of the perturbations can be stretched,
this hidden sector ``causal'' perturbation can contribute
significantly to the superhorizon (``acausal'') perturbations that are
required to explain the large scale structure of our universe and the
peaks in the microwave background \cite{boom}.  

Above we have focused on a 2 brane situation. However,
our results apply equally to a single brane.  For example, density
fluctuations produced in the bulk will have the same projection
on our brane (as from a hidden brane) 
and may thus be interesting for structure formation
or for providing constraints.

Even if our brane does
have an inflationary period to solve the flatness and the horizon
problem, lensed perturbations may be essential for structure
formation if the inflation itself produces perturbations that
are too small (e.g., if the height of the inflaton potential
is too low). 
%
%
In that case generation of perturbations elsewhere, such as on a hidden brane,
that are propagated to our brane in the way described in the paper,
may be essential to the formation of structure in the universe.
In fact perturbations generated in the bulk would carry the same
exponential suppression factors we have been describing.
However, for the lensed perturbations to serve this purpose, the
causal perturbations must be scale invariant.
We close this paper by presenting some ideas for future studies.

\subsection{Generation of Density Perturbations}

Although we leave the generation of scale invariant causal
perturbations to future studies, here we speculate about possible
origins of perturbations on the hidden brane.
One possible mechanism for the generation of scale invariant density
fluctuations on the hidden brane would be a second order phase
transition near criticality, where the order parameter field has scale
invariant two point correlations.  Another possible mechanism for
generation of scale invariant fluctuations is when a hidden brane has
a period of inflation, say for a few e-foldings, during which very
large perturbations, $\delta\rho_h /\rho_h \sim 1$ are produced.  While it
is not much harder to get a few efoldings than it is to get 60, the
advantage here is that the potential does not need to be fine-tuned to
be flat.  If this inflation is due to a potential, then it is the
height of the potential that determines the amplitude of the density
fluctuations. In ordinary inflationary scenarios, the ratio of the
height of the potential to the $({\rm width})^4$ must be very small,
less than $\sim (10^{-8})$ \cite{afg}; this small ratio usually
requires a fine-tuned small parameter.  On the hidden brane, on the
other hand, since the amplitude of the fluctuations can be as large as
we like, the height of the potential can be anything. Then the
potential does not need to be flat and there do not need to be any
small parameters.  For example, the width and height of the potential
may both be of order the Grand Unification scale $M_{GUT}$.  
If there are many branes sitting in the
warped extra dimensions, then as long as one of them generates large
perturbations e.g. by having a brief inflationary period, then we will
feel the existence of these perturbations and they will appear to us
to be much smaller in amplitude.

Of course, the flip side of this is that for this causally induced
acausal perturbation scheme to work, the scale invariance must not be
disturbed during the time structure formation commences.  In other
words, not only must the scale invariant perturbations be set up, it
must be protected until structure formation is well under way.  This
seems to be a significant model building challenge.  In particular,
the study of this issue will depend on how the transition (if any)
from the asymmetrically warped metric to a non-asymmetrically warped
metric takes place.

We have not investigated the time evolution of the lensed perturbations.
Our perturbation equations are valid at best over
a time interval much smaller than $1/H$, where Hubble constant $H$ is the
expansion rate of the universe.
A study of the time evolution of the perturbations would require 
more model dependent investigations by perturbing about
fully time dependent backgrounds.  Some work in braneworld density
perturbations in time dependent backgrounds can be
found in \cite{bruck,brandenberger,koyama,langlois,kodama,veneziano}.
We also have not investigated how 4D gravity will be recovered after
these perturbations are generated.  Because the Lorentz asymmetry in
warping needs to be strong, 4D effective theory description probably
will not be valid in describing the transition process.

\subsection{Flatness}

Here we briefly describe an attempt to solve the flatness problem due
to these lensed perturbations.  One statement of the flatness problem is
that there are fluctuations in density on our brane that give rise to
too much curvature.  One might hope to smooth out large fluctuations
on our brane by somehow ``averaging'' them with small fluctuations
appearing from some hidden brane(s).  As we will argue, one finds that
the amplitude of perturbations on our brane can be reduced by at most
$1/\sqrt{2}$ due to the effects from one hidden brane, or $1/\sqrt{N}$
where $N$ is the number of hidden branes.

Let us first consider a scenario with two branes: our observable brane
1 and a hidden brane 2.  We begin with a conserved stress tensor of a
perfect fluid on each of the branes.  When we perturb the fluid with
small fluctuations, we find that the system looks like a set of
coupled oscillator equations; the warping of the geometry plays a role
in that it enters in the mechanical tension.  One might hope that
small vibrations of brane 2 might absorb vibrations of brane 1, i.e.,
that small perturbations projected from the hidden brane might reduce
large perturbations on our observable brane.  However, we find that
the reduction of perturbations on our brane is not significant.  In
fact, equipartition of energy tells us that on the average the largest
amount of energy that can be lost by our system 1 is half of the
original energy. Thus vibrations of brane 1 can be reduced by at most
1/$\sqrt{2}$ on the average.

If there are $N$ hidden branes instead of just one, then equipartition
of energy implies that the amplitude of perturbations on our brane can
be reduced by at most $1/\sqrt{N}$.  In addition, our brane must be
sitting at a special place in the geometry, such that the direction of
lensing of the perturbations from the other branes makes them all
appear small to us.

Hence, even though a number of attempts have been made in the context
of extra dimensions to find an alternative mechanism for solving the
flatness problem \cite{flatness}, the usual inflationary scenario
seems most robust, thus far.  Ekpyrotic scenarios \cite{ekpyr} 
use the flatness of BPS branes. However, this approach begins with
a particular choice of initial conditions rather than providing
a dynamical explanation of the flatness of our universe.  (Still, of
course there may be quantum cosmological mechanisms that drive the
amplitude to be peaked in such configurations.)

On the other hand, braneworld extra
dimensions may make the flatness problem worse
\cite{isometrybreakingpaper2}.\footnote{For an example of a resolution, see
Ref. \cite{bae,moon}.} 
Ref.\cite{isometrybreakingpaper2} first pointed out 
a new signature for the existence of extra dimensions in the
context of braneworld scenarios:
due to the ``nonflatness'' of the bulk, gravitational waves may travel
at different speeds than photons from sources. (See
Ref. \cite{csaki} for a related perspective and Ref. \cite{kraus} for
earlier reference to this phenomenon.)  Then if one were to detect
gravitational waves and find no time difference between arrival times
of photons and gravitational waves, this detection would
present yet another aspect
of the cosmological flatness problem.


%


\subsection{Speculation on Dark Energy}

We end with one last comment.  We have shown that, in extra dimensions
with warped geometries, density perturbations appear to have different
amplitudes on our brane than where they were originally produced.
Hence, it is tempting to speculate that large energy densities with a
negative equation of state on the hidden sector brane may be projected
to a small energy density with a negative equation of state on our brane, 
thereby explaining the current acceleration of the universe indicated most
dramatically by supernova observations \cite{SN1,SN2}.  
An attempt to explain the acceleration in terms of an energy density
requires the energy density to be small. While the generation of such
a small value on our brane
typically requires fine-tuning, lensing of a larger
value generated elsewhere may suggest an improvement.
This question deserves further investigation.

\acknowledgments

We thank Csaba Csaki, Lisa Everett, Arthur Hebecker, Rocky Kolb,
Jianxin Lu, Antonio Riotto, Riccardo Rattazzi, Ren-Jie Zhang, and
particularly Robert Brandenberger, Jim Liu, and John March-Russell for
useful conversations.  We acknowledge support from the Department of
Energy via the University of Michigan.  K.F. thanks Columbia
University for hospitality during her visit.


\begin{appendix}
\section{Back reaction}

Based on our work in the body of the paper, one might naively expect a
reverse lensing effect: if large density perturbations on a hidden
brane appear to be small on our brane, then one might fear that small
density perturbations on our brane might appear larger and larger as
one goes out to large distances, eventually leading to back
reactions. However, we show here that, for point sources on our brane,
factors cancel each other out such that this is not the case. We find
agreement on this point with \cite{GKR}.  On the other hand, we show
that for non-pointlike sources, larger back reaction is expected.

Below we first consider a point source and then an extended source on
our brane; both are assumed to be static.  We ask the question: what
is the resulting gravitational potential at large distances?  We find
that 4D momentum near zero mode contributions are exponentially
enhanced but all higher momentum modes are suppressed.  For the point
source, when one adds up the contributions from all modes to invert
the Fourier transform, one finds that the number of exponentially
enhanced modes form an exponentially small set, and hence the sum is
not exponentially enhanced.  However, as the source becomes
increasingly delocalized (less point-like) the exponentially enhanced
zero mode contributions becomes increasingly important.  Hence, the back
reaction is increasingly significant for less localized density
perturbations.  This is related to the fact that for infinite
charge distributions of codimension one objects, the energy associated
with brining in a test charge from infinity is infinite.

Let us now show the mathematical details of this result.
 Eq. (\ref{eqorig}), the linearized field equation about the AdS
 metric, can be rewritten as
\begin{eqnarray*}
\frac{1}{\sqrt{g}}\partial _{a}[\sqrt{g}g^{ab}\partial _{b}]\tilde{h}_{rq} & = & \frac{1}{f}[\tilde{h}_{\mu r}^{\, \, \, ,\mu }(u_{i})+\tilde{h}_{\mu q\, \, \, ,r}^{\, \, \, ,\mu }(u_{i})-\tilde{h}(x,u_{i})_{,rq}\\
 &  & -\int _{u_{i}}^{u}du''\frac{R}{3M^{3}}T_{55,rq}+f'\eta _{rq}\frac{R}{6M^{3}}T_{55}+\\
 &  & -\frac{1}{M^{3}}(T_{qr}-\frac{1}{3}g_{qr}T^{s}_s)]
\end{eqnarray*}
 where \( f(u)\equiv e^{-2u/R} \), \( R \) is the radius of curvature
for the AdS space, $T_{qr}$ is the 5D stress energy tensor, and the
tildes indicate the perturbation is calculated in a particular gauge
in which fictitious bulk instabilities due to nonlocal sources are
removed. The first three terms on the right hand side correspond to
the initial conditions, and the terms involving \( T_{55} \) are gauge
dependent. The Green's function for this system has been calculated in
an earlier section of our paper in the form\[ G(x,u;x_{0},u_{0})=\int
\frac{d^{4}p}{(2\pi )^{4}}e^{ip\cdot (x-x_{0})}H(p,u,u_{0}).\]

Let us focus on the situation where a localized source on the Planck
brane seems to give a large back reaction far away in the bulk.
We found \[
H(p,L,0)=H_{>}(p,L,0)=\frac{be^{bL}}{p^{2}[I_{1}(e^{bL}\frac{p}{b})K_{1}(\frac{p}{b})-I_{1}(\frac{p}{b})K_{1}(e^{bL}\frac{p}{b})]}\]
where \( > \) symbol indicates that this Green's function is only
valid for the domain for which \[
L>0.\]
 
Now we take as our starting point: given a point source on the
brane at \(u=0\), what behaviour does one find at large distances?
We take the localized
source at \( u=0 \): i.e. \[
T_{00}\propto m\delta (u)\delta ^{(3)}(x-x_{0})\]
where \( m \) is the mass.
Accounting for the contribution only from
the sources localized to the brane in the bulk, we can write the associated
\( \tilde{h}_{00} \) as \begin{eqnarray}
\tilde{h}_{00}(x_{0},u=L) & = & c_{3}m\int \frac{d^{3}p}{(2\pi )^{3}}H_{>}(p,L,0)|_{p_{0}=0}\label{eq:approxpot} 
\end{eqnarray}
where \( c_{3} \) is a constant. Let us try to estimate this integral.
First, we expand \( p^{2}H_{>} \) to second order in \( p \) to
obtain\[
p^{2}H_{>}(p,L,0)=\frac{2be^{2bL}}{(e^{2bL}-1)}+p^{2}\left[ \frac{-e^{6bL}+4bLe^{4bL}+e^{2bL}}{4b(e^{2bL}-1)^{2}}\right] +O(p^{4})\]
Note that \begin{equation}
\label{eq:zeromomentumlimit}
\lim _{p\rightarrow 0}H_{>}(p,L,0)\sim \frac{2b}{p^{2}}.
\end{equation}
 We can expand this to leading order in \( e^{bL} \) to obtain\[
p^{2}H_{>}(p,L,0)=2b-p^{2}(\frac{e^{2bL}}{4b}-L+\frac{1}{2b})\]
From the fact that \( H_{>}(p,L,0) \) is positive definite and rapidly
falling, we can approximate Eq. (\ref{eq:approxpot}) by integrating
up to \[
p_{max}=\frac{2\sqrt{2}b}{\sqrt{e^{2bL}-4bL+2}},\]
resulting in\[
\tilde{h}_{00}(x_0,u=L)\approx \frac{8c_{3}mb^{2}}{6\pi ^{2}\sqrt{\frac{1}{2}e^{2bL}-2bL+1}}\approx \frac{8\sqrt{2}c_{3}mb^{2}e^{-bL}}{6\pi ^{2}}+O(e^{-2bL})\]
which indeed drops off as \( e^{-bL} \). We find that there is {\it no} 
large back reaction in this case. Note that this result has
been obtained earlier by Ref.\cite{GKR}.

Now that we have obtained this result, it remains for us to reconcile
it with our previous naive conclusion that density perturbations are
magnified by an exponential factor far away from the source.
Apparently, to reconcile our naive result with the exponentially
\emph{reduced} potential in the brane at \( u=L \) (in the previous
paragraph), one needs to introduce an effective energy density at \(
u=L \) that is exponentially \emph{larger.} Let us try to see why that
is.  For the mode with zero 4D momentum, we have the mode equation
\begin{equation}
\label{eq:zeromode5dequation}
\tilde{h}_{00,55}+2\frac{f'}{f}\tilde{h}_{00,5}=\frac{-1}{M^{3}f}T_{00}
\sim \frac{-\rho_1 \delta(u)}{M^3}
\end{equation}
 The friction term gives an enhancement factor for the \( \tilde{h}_{00} \)
contribution from the \( u \) dependence; i.e., if the source initially
is at \( u=0 \) and we are evaluating \( \tilde{h}_{00} \) at \( u=L \),
the source at \( u=0 \) will be amplified as it is propagated to
\( u=L. \) To see this, note that since \( T_{00} \) is localized
to \( u=0 \), the zero 4D momentum mode equation is solved
in the bulk by\[
\tilde{h}_{00}(k=0)=e^{4bu}j(x)\]
 for some function \( j(x) \) independent of \( u \). Hence, we
see that the total enhancement of the zero mode in propagating from
\( u=0 \) to \( u=L \) is \( e^{4bL}=1/f^{2}(L) \). Therefore,
there needs to be an enhancement of \( 1/f(L)^2 \) in the fictitious source
\( \underline{T_{qr}} \). This is precisely what we obtained in the
general propagator analysis in an earlier section.

However, as we saw in the analysis at the beginning of this Appendix,
the enhancement of the zero 4D momentum mode does not mean that the
source at \( u=0 \) is giving a large back reaction far away from
the source. The resolution of the apparent paradox is as follows.  The
configuration space \( \tilde{h}_{qr} \) need not be large even though
some Fourier amplitudes are;  the amplitude drops off rapidly
away from the zero momentum vector due to the fact that 
gravity is not confined
to the brane. To be more explicit, although Eq. (\ref{eq:zeromomentumlimit})
is exponentially larger than \( H_{>}(0,L,L) \), it has a phase space
of \( p_{max}^{3}\sim 10b^{3}e^{-3bL} \) which is exponentially suppressed
by a larger exponent than the exponential enhancement coming from
\( H_{>}(0,L \),0).  In other words, \( H_{>}(0,L,0)/H_{>}(0,L,L)=e^{4bL} \)
but \( \tilde{h}_{00}\propto \int \frac{d^{3}p}{(2\pi )^{3}}H_{>}(p,L,0)|_{p_{0}=0}\sim H_{>}(0,L,p_{max})p_{max}^{3}\propto e^{-bL} \)
. Hence, even with the Fourier transform of the zero 4D momentum mode
large, there is no necessarily large back reaction. The zero 4D momentum
mode having a large amplitude means that density perturbations on
length scales larger than \( b^{-1}e^{bL} \) do receive an exponential
enhancement.

We have shown above that, contrary to our naive expectations, 
point sources do not necessarily produce back reactions.  However,
it remains to consider extended sources.
Note that Eq. (\ref{eq:approxpot}) valid for a point-like source can
be easily generalised to a spatially smooth energy density:
\begin{equation}
\tilde{h}_{00}(x_{0},u=L)  =  \int \frac{d^{3}p}{(2\pi
  )^{3}}H_{>}(p,L,0)|_{p_{0}=0} \xi(x_0,\vec{p}) 
\end{equation}
where 
\begin{equation}
\xi(x,\vec{p})\propto e^{i \vec{p} \cdot \vec{x}} \int d^3 x'
\rho(\vec{x}') e^{-i \vec{p}\cdot \vec{x}'}
\end{equation}
and $\rho(\vec{x})$ is a smooth 4D energy distribution confined to the
brane at $u=0$.  Since we have seen that the exponential suppression
of $\tilde{h}_{00}$ derived from the small phase space integral about
$\vec{p}=0$, we see if $\xi$ is sufficiently peaked about $\vec{p}=0$
(delocalized source), we can naively obtain a large back reaction.  A
more careful treatment of this issue with the time dependence included
deserves further investigation.
\end{appendix}

\end{document}